\documentclass[12pt,a4paper]{article}
\usepackage{a4wide}
\usepackage{latexsym}
\usepackage{epsf}
\usepackage{amssymb}
\begin{document}
\def\be{\begin{equation}}
\def\ee{\end{equation}}
\def\bea{\begin{eqnarray}}
\def\eea{\end{eqnarray}}

\def\pd{\partial}
\def\a{\alpha}
\def\b{\beta}
\def\bi{\begin{itemize}}
\def\ei{\end{itemize}}
\def\g{\gamma}
\def\d{\delta}
\def\m{\mu}
\def\n{\nu}
\def \h{\mathcal{H}}
\def \hh{\mathcal{G}}
\def\t{\tau}
\def\p{\pi}
\def\th{\theta}
\def\l{\lambda}
\def\O{\Omega}
\def\r{\rho}
\def\cd{\cos{y_2}}
\def\cu{\cos{y_1}}
\def\su{\sin{y_1}}
\def\sd{\cos{y_2}}

\def\s{\sigma}
\def\e{\epsilon}
  \def\scri{\mathcal{J}}
\def\cM{\mathcal{M}}
\def\tcM{\tilde{\mathcal{M}}}
\def\RR{\mathbb{R}}

\hyphenation{re-pa-ra-me-tri-za-tion}
\hyphenation{trans-for-ma-tions}


\begin{flushright}
IFT-UAM/CSIC-04-01\\
hep-th/0401220\\
\end{flushright}

\vspace{1cm}

\begin{center}

{\bf\Large   Goursat's  Problem and the Holographic Principle.}

\vspace{.5cm}

{\bf Enrique \'Alvarez, Jorge Conde and Lorenzo Hern\'andez }

\vspace{.3cm}

\vskip 0.4cm  
 
{\it  Instituto de F\'{\i}sica Te\'orica UAM/CSIC, C-XVI,
and  Departamento de F\'{\i}sica Te\'orica, C-XI,\\
  Universidad Aut\'onoma de Madrid 
  E-28049-Madrid, Spain }

\vskip 0.2cm

\vskip 1cm

{\bf Abstract}

\end{center}

\begin{quote}

The whole idea of holography as put forward by Gerard 't Hooft assumes that data on
a boundary determine physics in the volume. This corresponds to a Dirichlet problem for 
euclidean signature, or to a Goursat (characteristic) problem in the lorentzian setting. 
Is this last aspect of the problem that is explored here for Ricci flat spaces with 
vanishing cosmological constant.
  
\end{quote}


\newpage

\setcounter{page}{1}
\setcounter{footnote}{1}
\newpage
\section{Introduction}
Ever since its inception,    the whole idea of 
{\em holography} (cf. \cite{'tHooft}
\cite{Susskind}) stand as one of the most original and mysterious suggestions ever 
made in fundamental physics.
\par 
From a certain abstract viewpoint,  it certainly includes at least two facts . 
\par
The first one is that it should
be possible  to recover volume information on the physical fields 
from data given on a certain
surface, namely the boundary of the volume. That is, fields in the volume do not 
obey an ordinary
Cauchy problem (for which also derivatives of the field at the boundary would also
 be needed),
but rather a degenerate one, in which the derivative cannot be imposed independently.
 This
 happens both for timelike and null initial surfaces.  
\par
Second, the volume symmetries (i.e. diffeomorphism
invariance) should guarantee conformal (or at least scale) invariance on the boundary. 
\par
In addition, although the main thrust of the holographic principle lies in its 
application to the fundamental degrees of freedom of quantum gravity, it seems
 sensible to assume that there is at least a regime in the space of parameters in which
these properties are true already at a classical level.
\par

Holography for vanishing cosmological constant remains quite mysterious to this day, and in any 
case is strongly suspected to be realized (if at all) in a subtler from than a 
Conformal Field Theory (CFT); 
indeed Witten (cf. \cite{Witten98}) coined the name {\em Structure X} to refer to it(cf.
also \cite{Arcioni}). 
\par

When the cosmological constant is negative, as in the conjectured duality 
of Maldacena ( \cite{Maldacena}), the mapping takes place because 
the conformal 
boundary in the Penrose sense (\cite{Penrose}) 
of Anti de Sitter (AdS) is timelike, so that data on this boundary determine the interior
dynamics, with some qualifications (\cite{Fefferman}\cite{Witten}). Curiously enough the 
ordinary Cauchy problem is not well posed in spaces of constant negative curvature, unless
some extra physical hypothesis are assumed (cf. the discussion in \cite{Avis}). Of course,
this rather vague ideas can be implemented in a precise way in the supersymmetric context
of $IIB$ strings in $AdS_5\times S_5$ with $N$ units of Ramond-Ramond (RR) flux. In this case, 
the conformal group is realized as an isometry group in the bulk, so that the boundary theory
must be conformal ($\beta=0$).
\par
Recently we have pointed out some indications of holographic behavior in a rather
general class of Ricci flat  string backgrounds  with vanishing cosmological constant
(\cite{Alvarez}) of the form
\be
M_{10}\equiv A_{6}\times C_{4}
\ee
where $C_4$ is an internal compact manifold, and $A_6$ will be denoted 
by the name {\em ambient space}. No RR backgrounds are excited,
so that this background is universal. In this ambient space lives a codimension two
{\em euclidean} four-manifold, which will be interpreted as the {\em spacetime} 
$M_4\subset A_6$.
The spacetime coordinates will be denoted by 
$y^i \equiv\vec{y}$  and its metric by $g_{ij}dy^i dy^j$;  whereas the extra two coordinates of 
the ambient 
space by $\rho\in\mathbb{R}^{+}$ 
and $T\in\mathbb{R}^{+}$, where $\rho$ is spacelike and $T$ timelike.
There is then a natural five-dimensional {\em boundary} defined in this patch by
\be
\pd A_6\equiv \{\rho=0\}
\ee

It is worth remarking that this boundary is {\em null}, and it is then essentially 
equivalent to 
{\em Penrose's conformal boundary}, which is known to play a central r\^ole in holography.
\par
An important fact worth remembering is that the {\em conformal boundary} of a 
(conformal) boundary does not necessarily vanish.
\par
There are then two conformal boundaries in this setting: the finite one 
at $\rho=0$, which is also a mathematical boundary located at finite distance, and 
the usual conformal boundary at infinity, the appropiate $\mathcal{J}^{+}$.
\par
It will become clear in the sequel that in many cases we will be able 
to interpret the whole space as the {\em curved interior} of some 
{\em light-cone}, which itself represents the boundary.

\par
Those spaces (from now on we shall work in arbitrary dimension $d=n+2$, because most
results are quite general) enjoy a {\em homothecy} that is, a conformal Killing vector (CKV)
 with constant 
conformal factor, which acts on the metric through a scale transformation
\be
\pounds(k)g_{\a\b}=2 g_{\a\b}
\ee 
Canonical coordinates can be chosen such that the whole ambient space metric  reads:
\be\label{ambiente}
ds^2=\frac{T^2}{l^2} ds^2(y,\rho) +\rho dT^2 + T d\rho dT
\ee
The CKV itself is then related to the preferred timelike coordinate
through
\be\label{ckv}
k\equiv T\,\frac{\pd}{\pd T}
\ee
where the metric reduces on the null codimension-one hypersurface $\rho=0$
 to the (riemannian) $n-$dimensional spacetime metric (up  to a rescaling):
\be\label{frontera}
ds^2(y,\rho=0) = \frac{T^2}{l^2} g_{ij}(y) dy^i dy^j.
\ee
\par
The norm of the CKV is given by
\be
k^2= \rho T^2
\ee
The boundary then has an invariant characterization as a {\em Killing horizon}, the set of
 points
where the norm of the CKV vanishes. 
\par
Under Weyl rescalings of the metric (\ref{ambiente}),
\be
g_{\a\b}\rightarrow \hat{g}_{\a\b}\equiv\Omega^2 (T,\rho,y) g_{\a\b}
\ee
the same vector $k$ given by (\ref{ckv}) remains a CKV, because
\be
\pounds(k)\hat{g}_{\m\n}=\bigg(2+\frac{\pd\log{\Omega}}{\pd\log{T}}\bigg)\hat{g}_{\m\n}
\ee
and its norm changes by a conformal factor
\be
k^2=\Omega^2 \rho T^2
\ee
This means that the whole setup is {\em conformally invariant}. In particular, the 
interpretation of the boundary as the
Killing horizon of the CKV survives Weyl rescalings. 
\par
The purpose of the present paper is to study the interplay bulk/boundary in this framework.
One of the main characteristics is the fact that the {\em finite boundary} (that is, the 
one at $\rho=0$)
is precisely located at {\em finite distance}, in spite of being also a conformal boundary;
as we have already said there is always in addition the boundary at infinity , 
$\mathcal{J}^{+}$ as 
well, which is infinitely far away as usual. We shall mainly be concerned with the 
appropiate generalizations  of the bulk-boundary Green functions, as well as with the 
symmetries of the finite boundary action.  In this paper we assume that all relevant 
curvatures are
small enough so that a (super)gravity treatment is a useful first approximation.

\section{The simplest example: the Milne Universe}
The whole idea of the present approach implicity assumes that the ordinary Cauchy problem
has been replaced by a characteristic one (cf. \cite{Courant}), usually called {\em Goursat's
problem} in the mathematical literature.
In the former,  Cauchy data on a spacelike surface (such as $t=const$ in flat space)
are given. This means, for the wave equation, giving the field and the normal (time) derivative
of it in the initial surface, and the solution is then fully determined in the causal
development of the Cauchy surface.
The characteristic problem, on the other hand, specifies half of the data (i.e.,
the field itself) on a characteristic surface of the hyperbolic equation, such as the 
light cones in the flat case. The solution is then fully determined in the 
inside of the cone only. Curiously enough, if we consider the inside of the light cone 
as a mildly singular manifold
the ordinary Cauchy problem is delicate.
\par
As a simple example, where however, all the ideas get illustrated, let us think for a moment 
on the forward light cones $N_{+}$ on flat (n+2)-dimensional 
space, the simplest possible background. The inside of the forward light cone is what is usually
denoted by Milne's Universe (\cite{Birrell}). The general situation can be studied with few
modifications .
\par
There are then two boundaries in our space (cf. Figure). One is the finite conformal boundary,
i.e., the future light cone of the origin, $N_{+}(0)$. The other one is the future null 
infinity, $\mathcal{J}^{+}$. The Penrose diagram of the Milne universe corresponds 
 thus to the portion of Minkowski space shown in the figure.
\par

\begin{figure}[!ht] 
\begin{center} 
\leavevmode 
\epsfxsize= 6cm

\epsffile{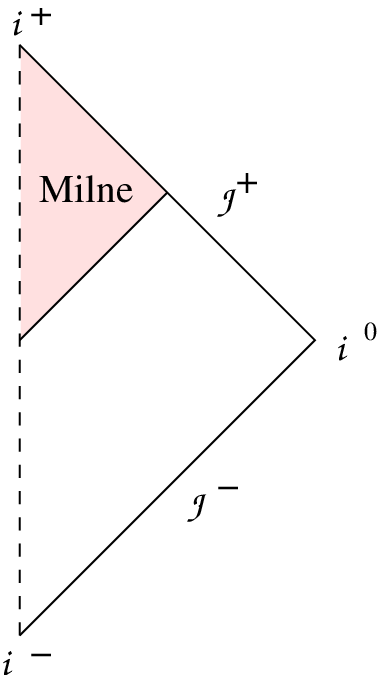} 

\end{center} 
\end{figure}
\par

The starting point is then the flat metric
\be
ds^2=\eta_{\m\n}dx^{\m}dx^{\n}=(dx^0)^2-d\vec{x}^2
\ee
($x^i=(x^1\ldots x^{n+1})$ and we employ $x^0=t$ indistinctly).
The equation 
of the light cones is 
\be
x_0=r \equiv|\vec{x}|
\ee

conveying the fact that they are null surfaces themselves, since their normal vector
\be
n\equiv(1,n^i\equiv\frac{x^i}{r})
\ee
is null.
By the way,these coordinates have nothing to do with 
the canonical coordinates  introduced above. The explicit relationship is
\bea\label{cartesian}
&&T=x_{-}\equiv x_0-x_{n+1}\nonumber\\
&&y^i=\frac{l x^i}{x_{-}}(i\neq n+1)\nonumber\\
&&\rho=\frac{x_{\m}x^{\m}}{x_{-}^2}
\eea

 Their local structure is $S^{n}\times\mathbb{R}^{+}$, and a point in $N_{+}$
can be specified by $(x_0,n^i)$, where $x_0\in \mathbb{R}^{+}$ and $\vec{n}^2=1$ is a point
on the unit n-dimensional sphere, $S_n$, that is, a $(n+1)$-dimensional structure. 
The light cone can be visualized as a $S_{n}$ sphere of radius $x_0$.
\par

The induced metric  is, however, 
degenerate (that is, as a matrix it has rank $n$), because
the time differential is totally absent from the line element:
\be\label{roll}
ds_{+}^2=x_0^2 d\Omega_n^2
\ee
where $d\Omega_n^2$ is the metric on the unit n-sphere, $S_n$, which in terms of
angular variables reads: 
\be\label{cono}
d\Omega_n^2\equiv d\theta_n^2 + \sin{\theta_n}^2 d\theta_{n-1}^2+\ldots + \sin{\theta_n}^2
\sin{\theta_{n-1}}^2\ldots \sin{\theta_2}^2 d\theta_1^2
\ee
This means that, although singular as a metric on $N_{+}$, the metric is perfectly regular 
(actually
the standard one) as a metric on the n-spheres $t=constant$.
\par

The invariant volume element, however, vanishes, due to the fact that the determinant of 
the induced metric is zero.
\par
The metric con $N_{+}$ inherits the homothecy giving rise to scale transformations on the metric
which now reads $k\equiv x_0\pd_{x_0}$ and still obeys
\be
\pounds(k)g_{\a\b}=2 g_{\a\b}
\ee

\par

Remarkably enough, the complete set of isometries of the 
{\em three}-dimensional metric (\ref{roll}) includes the full Lorentz group,
$SO(1,3)$.  Please note that isometries are well-defined, even for singular metrics, 
through the vanishing Lie-derivative condition
$\pounds (k) g_{\m\n}=0$, reflecting the invariance of the metric under the corresponding 
one-parametric group 
of diffeomorphisms, although of course this is not equivalent to $\nabla_{\m} 
k_{\n}+\nabla_{\n} k_{\m}=0$ because the 
covariant derivative (that is, the Christoffel symbols ) is not well defined owing to 
the absence of the inverse metric.
\par
The six Killing vectors that generate $SO(1,3)$ are : 
\bea
&&J_1=\cos{\phi}\frac{\pd}{\pd\theta}-\cot{\theta}\sin{\phi}\frac{\pd}{\pd \phi}\nonumber\\
&&J_2=\sin{\phi}\frac{\pd}{\pd\theta}+\cot{\theta}\cos{\phi}\frac{\pd}{\pd \phi}\nonumber\\
&&J_3=\frac{\pd}{\pd \phi}\nonumber\\
&&K_1=-x^0 \sin{\theta}\sin{\phi}\frac{\pd}{\pd x^0}-\cos{\theta}\sin{\phi}\frac{\pd}{\pd \theta}
-\frac{\cos{\phi}}{\sin{\theta}}\frac{\pd}{\pd \phi}\nonumber\\
&&K_2=x^0 \sin{\theta}\cos{\phi}\frac{\pd}{\pd x^0}+\cos{\theta}\cos{\phi}\frac{\pd}{\pd \theta}
-\frac{\sin{\phi}}{\sin{\theta}}\frac{\pd}{\pd \phi}\nonumber\\
&&K_3=x^0 \cos{\theta}\frac{\pd}{\pd x^0}-\sin{\theta}\frac{\pd}{\pd \theta}
\eea
Let us point out, however, that a slight shift in viewpoint uncovers an infinite group of 
isometries.
\par
The light cone can indeed be considered as the infinite curvature limit of EAdS. 
(cf. \cite{AlvarezV}). 
The exact relationship between cartesian and horospheric coordinates in the 
infinite curvature limit is:
\bea
&&x_0=\frac{1}{2 z}(y_T^2+1)\nonumber\\
&&x_3=\frac{1}{2 z}(y_T^2-1)\nonumber\\
&&x_T=\frac{y_T}{z}
\eea
where the subscript {\em transverse} refers to the $(1,2)$ labels: $y_T\equiv (y_1,y_2)$. It is worth pointing out that
the coordinate $z$ has got dimensions of energy, whereas the $y_T$ are dimensionless.
\par
Horospheric coordinates then break down when $x_0=x_3$; that is, when $z=\infty$. The metric
of the cone reads now
\be
ds^2=\frac{d y_1^2 + d y_2^2}{z^2}
\ee
\par
It is a simple matter to recover the Killings corresponding to the Lorentz subgroup.

But there are more Killing vectors. First of all, the two translational ones, 
obvious in these coordinates:
\bea
&&P_1\equiv\frac{\pd}{\pd y_1}\nonumber\\
&&P_2\equiv\frac{\pd}{\pd y_2}\nonumber\\
\eea
and some others as well. The fact that there is translation invariance in horospheric 
coordinates in $N_{+}$ is of great importance in the definition itself of the Green functions.
\par
It is actually possible to give the general solution of the Killing equation in closed form 
using our horospheric 
coordinates. Given an arbitrary {\em harmonic} function of the two variables $(y_1,y_2)$, it
 is given by:
\be
k\equiv (\frac{\pd^2}{\pd y_1^2} f)z\frac{\pd}{\pd z}+(\frac{\pd}{\pd y_1} f) \frac{\pd}{\pd y_1}
-(\frac{\pd}{\pd y_2}f) \frac{\pd}{\pd y_2}
\ee

The finite transformations corresponding to those Killing vectors are:
\bea
&&y_1\rightarrow y_1 + \pd_1 f\nonumber\\
&&y_2\rightarrow y_2 - \pd_2 f\nonumber\\
&&z\rightarrow \sqrt{(1+\pd_1\pd_1 f)^2 +(\pd_1\pd_2 f)^2}
\eea
The composition of two succesive transformations characterized by the harmonic functions 
$f$ and $g$ is equivalent to the function
\be
F(f,g)\equiv f(y_1,y_2)+g(y_1+\pd_1 f,y_2-\pd_2 f)
\ee
The conmutator function is then easily found to be
\be
[f,g]=f(y_1,y_2)+g(y_1+\pd_1 f,y_2-\pd_2 f)-g(y_1,y_2)-f(y_1+\pd_1 g,y_2-\pd_2 g)
\ee

\par
It is now clear that the isometry group of the four-dimensional light cone $N_{+}$ is an 
infinite dimensional group, 
which includes the Lorentz group as a subgroup.
\par
We find this to be a remarkable situation.
\par
Even more remarkable is the fact that in higher dimension, when the total space gets dimension
$d$, say, so that the light cone has dimension $d-1$, and in horospheric coordinates is
 characterized by $z$ and $\vec{y}\in\mathbb{R}^{d-2}$, in such a way that the metric reads
\be
ds^2=\frac{d\vec{y}^2}{z^2},
\ee
and the Killing equations are
equivalent to
\be\label{conf}
\pd_i k_j+\pd_j k_i = 2\d_{ij}\xi(\vec{y})
\ee 
for the total vector
\be
k=z \xi(\vec{y}) \pd_z + \sum_{i=1}^{d-2} k^i\pd_i
\ee
But the equations (\ref{conf}) are precisely the equations for the conformal Kiling
vectors of flat $(d-2)$-dimensional space, known to generate the euclidean 
conformal group, $SO(1,d-1)$, isomorphic to the $d$-dimensional Lorentz group. 
To be specific (\cite{Erdmenger}),
\be
\xi(\vec{y})\equiv \lambda-2 \vec{b}.\vec{y}
\ee
and the components on the $y$-directions read:
\be
k_i=a_i+\omega_{ij} y^j+\lambda y_i + b_i y^2-2 \vec{b}.\vec{y} y_i
\ee

representing translations ($a$), rotations ($\omega_{(ij)}=0$), scale transformations, 
$(\l)$, and 
special conformal transformations ($b)$.

\par
To summarize, the isometry group of the light cone at the origin, $N_{+}(0)$, is generically
the spacetime Lorentz group {\em except} in the four dimensional case, in which it expands
to the infinite group  we derived above.

\par
Also interesting are those transformations that leave invariant the metric up to a Weyl 
rescaling (which should include our group as a subgroup). Those are
the conformal isometries which in four dimensions span the so called the Newman-Unti (NU) 
group (cf. \cite{Penrose}), i.e.
\bea
&& x^0\rightarrow F(x^0,z,\bar{z})\nonumber\\
&& z\rightarrow \frac{a z + b}{c z + d}
\eea
where $z$ is the complex stereographic coordinate of the sphere $S_2$, and {\em not} the
horospheric coordinate.
The NU group is also an infinite dimensional extension of the M\"obius group. In the appendix 
we have worked out some illustrative examples. 
\par
The Bondi-Metzner-Sachs (BMS) subgroup
consists on those transformations which are linear in $x^0$.
\section{Riesz' potential in the massless case.}

The characteristic problem is much less well-known than the corresponding Cauchy problem. It only makes sense for hyperbolic equations, and then the problem is
to determine the solution in a suitable domain of dependence, given the field
in a characteristic surface (such as a light cone for the wave or Klein-Gordon 
equation). Is equivalent to a degenerate Cauchy problem in that the derivative
cannot be prescribed arbitrarily; the relationship with the Dirichlet problem
of the elliptic {\em euclidean} equation is subtle, and will be dealt with (through
some elementary examples)
in an Appendix.
It seems to have been completely solved for the wave equation by d'Adhemar in 1905 (cf.\cite
{Dadhemar}). Indeed, in the book \cite{Penrose} reference is made to the Kirkchhoff-d'Adhemar
formula. In general the integrals needed when the classic techniques of Kirkchhoff and Volterra
are directly applied are divergent. In one of the last chapters of the classic 
age of mathematical
physics, J. Hadamard introduced the concept
of {\em partie finie} \cite{Hadamard}  in order to give 
a precise recipe to compute them. 
We shall employ here, however, mainly the equivalent (although somewhat more general) 
alternative solution
elaborated by Riesz \cite{Riesz}, and based on analytically 
continuing integrals depending on a complex 
parameter in such a way that they are
convergent in a particular region of the complex plane, a method which was to become 
popular among physicists many decades later. The mathematical problems 
encountered are not  unrelated to the ones appearing when a precise meaning 
is given to the
equations of motion in general relativity, or even in classical electrodynamics.
Dimensional regularization (\cite{Blanchet}) can most 
likely be
 employed here as well although we shall not pursue this avenue in this paper.
\subsection{Generalities on the characteristic problem}
In order to have a first look at the main differences between Goursat's and Cauchy's problems,
let us  consider a scalar field in the interior of the light cone, $C_{+}$, with 
prescribed values on the cone itself, $N_{+}$ (all this in flat n-dimensional Minkowskian space)
\be
\Phi|_{N_{+}}(t=r,\vec{n})=\psi(t,\vec{n})
\ee
In ingoing (u) and outgoing (v) null coordinates
\bea
&&u\equiv t+ r\nonumber\\
&&v=t-r
\eea
the metric reads
\be
ds^2= du dv - \frac{(u-v)^2}{4}d\Omega_{n-2}^2
\ee
The boundary is now $v=0$, and it can be said that
\be
\Phi|_{N_{+}}(u,v=0,\vec{n})=\psi(u,\vec{n})
\ee
\par
It is plain that the $\frac{\pd}{\pd u}$ derivative is determined by the boundary condition:
\be
\pd_{u}\Phi(u,v=0,\vec{n})=\pd_{u}\psi(u,\vec{n})
\ee
but the $\frac{\pd}{\pd v}$ derivative instead is unknown in principle:
\be
\pd_{v}\Phi(u,v=0,\vec{n})\equiv f(u,\vec{n})
\ee
This is precisely the would-be extra data in a Cauchy problem, the analogous of the normal 
derivative of the field at the initial surface.
\par
In our case, however, this function $f$ is not arbitrary, but instead it is fully 
determined in terms of $\psi$ through the wave equation in the cone:
\be
4\pd_{u} f-\frac{2n-4}{u}(\pd_{u}\psi -f)-\frac{4}{u^2}\Delta_{\vec{n}}\psi=0
\ee
(where $\Delta_{\vec{n}}$ is the laplacian on the sphere $S_{n-2}$). The ordinary differential
equation that the unknown function $f$ obeys is easily solved:
\be
f(u,\vec{n})=\frac{1}{4}u^{-(n-2)/2}\int^{u}\tau^{(n-2)/2}d\tau \big(\frac{2n-4}{\tau}
\pd_{u}\psi(\tau)+\frac{4}{\tau^2}\Delta_{\vec{n}}\psi(\tau)
\big)
\ee

\par
Please note that
\be
\pd_t \Phi\big|_{t=r}=\pd_u\psi(u,\vec{n}) + f(u,\vec{n})
\ee
whereas
\be
\pd_r \Phi\big|_{t=r}=\pd_u\psi(u,\vec{n}) - f(u,\vec{n})
\ee
which are {\em different} in general.
\subsection{The Generalized Potential of Order $a$.}

Let us consider the lorentzian generalization of the Riemann-Liouville integral
(cf. Appendix)
\be\label{rl}
I^a f(x)=\frac{1}{H_n(a)}\int_{D^x_S} f(y) \tau_{xy}^{a-n}dy
\ee
where $\tau_{xy}$ is the proper time (geodesic distance) between the points $x$ and $y$, and 
\be
H_n(a)\equiv \pi^{n/2 -1}2^{a-1}\Gamma(a/2)\Gamma(\frac{a+2-n}{2})
\ee
and $D^x_S$ is the region bounded by the past light cone of the point $x$, $N_{-}(x)$, and the
codimension one hypersurface $S$.
This construct satisfies
\bea
&&I^a I^b= I^{a+b}\nonumber\\
&&\Box I^{a+2}=I^a
\eea
besides the characteristic property of the Riemann-Liouville integral
\be
I^0 f= f
\ee
One can then say in a certain sense that
\be
I^2 =\Box^{-1}
\ee

\subsection{Stokes'theorem at work}
Let us indeed consider the integral over an arbitrary n-dimensional chain $D$
\be
I\equiv\int_{D} f d*d g- g d*d f=\int_{D}d(vol)(f\Box g- g \Box f)
\ee
where the invariant volume element is defined by
\be
d(vol)\equiv \sqrt{|g|}dx^1\wedge\ldots\wedge dx^n
\ee
Stokes theorem guarantees that
\be
I=\int_{D}d(f*dg - g*df)+dg\wedge*df-df\wedge*dg=\int_{\pd D}(f*dg - g*df)
\ee
Let us now consider as $D$ the interior region bounded by two light cones in 
n-dimensional Minkowski space, one corresponding to the future of the 
origin $(N_{+}(0)$ which is the characteristic 
surface), and another one the past light cone of an arbitrary point,$(N_{-}(P)$,
where $P\equiv x^{\m}$.

\begin{figure}[!ht] 
\begin{center} 
\leavevmode 
\epsfxsize= 5cm

\epsffile{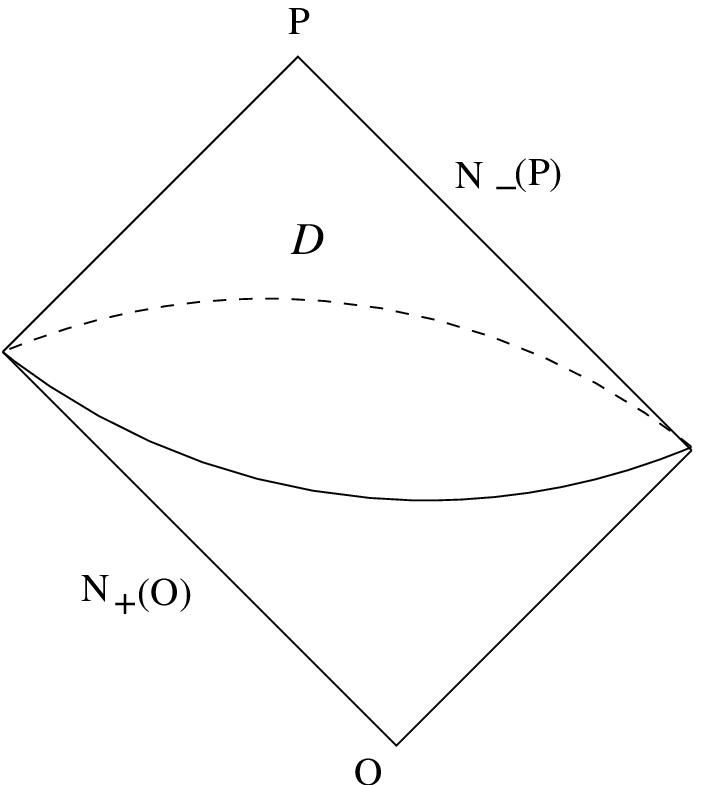} 

\caption{{\em }}
\label{fig:cero} 
\end{center} 
\end{figure}
\par

Let us now consider the functions $g\equiv \tau_{xy}^{2\lambda}\equiv(x-y)^{2\lambda}$ and 
$f\equiv\phi(x)$ is the
solution we are seeking for,
$\Box\phi(x)=0$ and $\phi|_{N_{+}(0)}=\psi$.
We then have
\be
2\lambda(n+2\lambda-2)\int_{D}\phi(y)(x-y)^{2\lambda-2}=\int_{\pd D} \left(\phi *d \tau_{xy}^{2\lambda}
 - \tau_{xy}^{2\lambda}*d\phi \right)  
\ee

The first integral is easily seen to be proportional 
 to the Riemann-Liouville integral 
\be
I^{2\lambda+n-2}\phi=\frac{1}{H_n(n+2\lambda)}\int_{\pd D} \left( \phi * 
d\tau_{xy}^{2\lambda}-\tau_{xy}^{2\lambda}*d\phi \right)
\ee
(where the well-known identity $z\Gamma(z)=\Gamma(1+z)$ has been used).
\par
The main thrust of Riesz approach comes here: the first member tends to $I^0 \phi=\phi$
(the only instance allowed by dimensional considerations) when
\be
\lambda\rightarrow 1- n/2
\ee
On the other hand, if we agree in {\em defining} the second member trough analytic continuation,
it is cleat that for $Re\, \lambda$ large enough the part of the boundary including $N_{-}(P)$
will not contribute, and we are left with the only task of 
computing the integral over $N_{+}(0)$.

\be\label{basica}
\phi(x)=(I^{0}\phi)(x)=\lim_{\lambda\rightarrow 1-\frac{n}{2}}\frac{1}{2^{n+2\lambda-1}\pi^{\frac{n}{2} -1} 
\Gamma(\lambda+\frac{n}{2})\Gamma(\lambda+1)}
\int_{y\in N_{+}(0)} {\left(\phi(y) * 
d\tau_{xy}^{2\lambda}-\tau_{xy}^{2\lambda}*d\phi(y) \right)}
\ee
\subsection{The concept of characteristic propagator}
The basic equation (\ref{basica}) , after integration by parts, 
can be expressed as an integral operator of the type
\be
\phi(x)=\int_{y\in N_{+}(0)} D_c (x,y) \psi(y)
\ee
mapping the characteristic data $\psi$ into the bulk solution $\phi$. We shall see in due
time that this definition is delicate, because there are cancellations between zeroes and poles
which appear only after integration.
\par
Some general properties can be however drawn from the general formula (\ref{basica}).
Please note for example that given the fact that if $x\in N_{+}$ $x+a$ does 
not necessarily belong to $N_{+}$,
there is no general reason why the propagator $D_c$ ought to depend on the difference $x-y$ only.
 That is, if for a general function $f$ we call $(T_a f)(x)\equiv f(x+a)$ then in order to 
show that the image of $T_a\psi$ is $T_a\phi$ we need translation invariance. 
\par
There is more in this point than meets the eye, as shall be seen in due course.
\par
Exactly the same argument shows that in our case, the image of the scaled function
$(T_{\Omega}\psi)(x)\equiv
\psi(\Omega x)$ is  $(T_{\Omega}\phi)(x)$, owing to the fact that acting on the first argument
\be
D_c(\Omega x,y)=\Omega^{2\lambda} D_c(x,y)
\ee

\subsection{Three and four-dimensional examples}

One on the main properties of the solution to Goursat's problem (shared with Cauchy's problem)
is that for even total spacetime dimensions the solution depends only on data defined on the
intersection of the past light cone of the point at which the solution is evaluated, with the
characteristic. This is the essence of {\em Huyghens'principle}, and we shall call the 
corresponding codimension two surface {\em Huyghens surface} ${\cal H}\equiv N_{-}(P)
\cap N_{+}(0)$. For odd spacetime dimensions
this is not so, and the integrals span the whole codimension one region of the characteristic 
preceding the 
Huyghens surface. 
\par

Let us be specific in the simplest situations. In a $n=2+1$ dimensional
spacetime
\be
*d\phi=\frac{\pd\phi}{\pd t}r dr\wedge d\theta+\frac{\pd\phi}{\pd r}r dt\wedge d\theta-
\frac{\pd\phi}{\pd \theta}r dt\wedge dr
\ee
On the future light cone of the origin, $t=r$ the last term vanishes, yielding ($u=t+r$)
\be
*d\phi= 2r\frac{\pd\phi}{\pd u}dr\wedge d\theta
\ee
Instead of using the coordinate $r$, it is convenient to employ (only to the effect of the
 computation
of the integral) the square of the proper time measured from the point 
$P\equiv (\bar{x}^{\m})$, that is
\be
R\equiv\tau^2= (\bar{x}-x)^2=\bar{x}^2+ 2 r(\bar{x}_1 \cos{\theta}+\bar{x}_2\sin{\theta}-
\bar{x}_0)
\ee
This leaves us with
\bea
&&\phi^{(3)}(\bar{x})=\lim_{\lambda\rightarrow -1/2}\int \frac{d\theta}{2\pi}\int_0^{\bar{x}^2}\Big[
\psi(R_{\bar{x}},\theta)\lambda R_{\bar{x}}^{\lambda-1}
\frac{R_{\bar{x}} - \bar{x}^2}{2(\bar{x}_1 \cos{\theta}+\bar{x}_2\sin{\theta}-\bar{x}_0)}
\nonumber\\
&&-R_{\bar{x}}^{\lambda}\frac{\pd\psi}{\pd u}\frac{R_{\bar{x}} - \bar{x}^2}
{2(\bar{x}_1 \cos{\theta}+ \bar{x}_2\sin{\theta}-\bar{x}_0)^2} \Big]dR_{\bar{x}}
\eea
Now we can write (we shall suppress the subindex in $R$ unless the equation is prone
to confusion)
\be
\frac{\pd\psi}{\pd u}=\frac{\pd\psi}{\pd R}(\bar{x}_1 \cos{\theta}+\bar{x}_2\sin{\theta}-
\bar{x}_0)
\ee
and performing integration by parts we can reach an expression without derivatives of $\psi$, which for our present purposes is best suited;
\be\label{pr}
\phi^{(3)}(\bar{x})=\lim_{\lambda\rightarrow -1/2}\int \frac{d\theta}{2\pi}\frac{1}{2
(\bar{x}_1 \cos{\theta}+\bar{x}_2\sin{\theta}-\bar{x}_0)}\int_0^{\bar{x}^2}dR
R^{\lambda-1}[(1 + 2 \lambda)R-2 \bar{x}^2\lambda]\psi(R,\theta)
\ee
\par

It is not difficult to check that when the data on the cone 
are constant, such as 
$\psi=1$ then the result of the integration is also constant, i.e.,
$\phi=1$.
\par
The solution thus obtained enjoys simple properties under rescalings; to the seed
\be
\psi_{\Omega}(\bar{x})
\ee
it corresponds the solution
\be
\Omega^{2 \l +1} \phi(\bar{x})
\ee
and the whole characteristic problem is scale invariant for the critical value of $\l$, as it should.
\par
By using a delta source as initial function
\be
\psi=\d (R-R^{\prime})\d(\theta-\theta^{\prime})
\ee
(which depends also on $\bar{x}$ through the variable $R$), we get the {\em characteristic boundary-bulk propagator}
\be
K_{(3)}^{b-B}(\bar{x},R^{\prime},\theta^{\prime})=\frac{H(\bar{x}^2-R^{\prime})H(R^{\prime})}
{4\pi (1+\lambda)}
\frac{[(1+2\lambda)R^{\prime}-2 \bar{x}^2 \lambda](R^{\prime})^{\lambda-1}}
{\bar{x}_1 \cos{\theta^{\prime}}+
\bar{x}_2 \sin{\theta^{\prime}}-\bar{x}_0}
\ee
where $H(x)$ is the Heaviside step function. 
It is instructive to consider the limit when the point $\bar{x}$ approaches $N_{+}(0)$, the future
light cone of the origin. Let us put
\be
\bar{x}^2=\epsilon^2
\ee
so that
\be
\bar{t}=\bar{r}+\frac{\e^2}{2\bar{r}}
\ee
An easy integration shows that
\be
H(\e^2-R^{\prime})H(R^{\prime})\sim \e^2\d(\e^2-R^{\prime})+o(\e^4)
\ee

The $\e$-dependent part of the integrand is then
\be
-\frac{\e^{2\lambda+2}}{[\bar{r}(1-\cos{(\theta^{\prime}-\bar{\theta})})+\frac{\e^2}
{2\bar{r}}]}
\ee
Now using the well-known fact that
\be
\lim_{\e\rightarrow 0}\frac{\e^{\b}}{(\e^2 +\vec{x}^2)^{\a}}=\frac{\pi^{\a- \b/2}\Gamma(\b/2)}{
\Gamma(\a)}\d^{n=(2\a-\b)}(\vec{x})
\ee
we get
\be
K_3\rightarrow \frac{2\bar{r}}{2\pi }\frac{\pi^{-\l}\Gamma(1/2)}
{\Gamma(1)} \d^{-2\l}(2\bar{r}\sqrt{(1-\cos{(\theta^{\prime}-\bar{\theta})}})
\d(R^{\prime})
\ee

In the limit when $\l\rightarrow -1/ 2$ this yields
\be
\frac{2\bar{r}\sqrt{\pi}\sqrt{\pi}}{2\pi}\frac{1}{\bar{r}}\d(\theta^{\prime}-\bar{\theta})
\d(R^{\prime})=\d(\theta^{\prime}-\bar{\theta})
\d(R^{\prime})
\ee

Let us now turn our attention towards the solution of the four-dimensional ($1+3$) 
characteristic problem, which can be
given  through (where we represent by $\omega\equiv(\theta,\phi)$ the polar coordinates of 
a point $\vec{n}$, $\vec{n}^2=1$ in the unit
two-sphere $S_2$, and $d\Omega_2\equiv \sin{\theta}d\theta\wedge d\phi$),

\be
\phi^{(4)}(\bar{t},\bar{r},\bar{\Omega})= \lim_{\lambda \rightarrow -1}\frac{1}{H_4} \int
d\Omega \int_0 ^{\frac{\tau^2}{2\h}} dt (2t)\left[\tau^2 +
2(\lambda+1) \h t\right]^{\l-1}\left[\tau^2-2(\l+1) \h t\right]  \psi(t, \Omega)
\ee

with the boundary value given by:

\be
\psi(t,\Omega) \equiv \phi^{(4)} (t,t,\Omega)
\ee
and the constant
\be
H_4 = \pi 2^{3+2\l}\Gamma(2+\l)\Gamma(1+\l) =\frac{2 \pi}{\l+1} 
\ee

We have liberally introduced Riesz' parameter $\l$ (eventually to be taken towards 
$\l\rightarrow -1$) 
into several definitions:

\be
\begin{array}{l}
\tau ^2 \equiv \bar{t}^{\ 2} -\l^2 \bar{r}^{\ 2}\\ 
\h \equiv \bar{t} + \bar{r}\l \cos{\g}\\ 
\cos{\g} \equiv \vec{\bar{n}}.\vec{n}
\end{array}
\ee
Again, it is not difficult to check that under rescalings
\be
\psi_{\Omega}\rightarrow \Omega^{2+2\l} \phi
\ee 
The characteristic boundary-bulk propagator can be easily read from the above formula

\be
K^{b-B}_{(4)}(\bar{t},\bar{r},\bar{\Omega} ; t,\Omega) = \lim_{\l \rightarrow -1} -\frac{\l+1}{\pi}t\left[\tau^2
-2\h t\right]^{\l-1}\left[\tau^2-2(\l+1) \h t\right] H(\frac{\tau^2}{2\h}-t)
\ee
We define an angular function $\hh$ through:
\be
\hh = 1+\l \cos{\g}
\ee
so that the kernel can be particularized to $N_{+}(0)$, yielding
\be \label{propa}
K|_{N_{+}} =\lim_{\l \rightarrow -1} -\frac{(\l+1)^2}{\pi}t't^{\l}\left[
t(1-\l^2)-2t'\hh\right]^{\l-1}\left[ t(1-\l)-2t'\hh \right]
\ee 
which can be easily shown to enjoy the properties of a delta-function when $\l\rightarrow -1$,
as it should.
\par

\section{The Characteristic boundary-bulk propagator for Masssive fields. }
In this case Riesz has shown \cite{Riesz} that the r\^ole of the proper time, $\tau^{a-n}$ in the 
Riemann-Liouville integral
is played by
\be
w^a(x,y)\equiv \frac{(\tau_{xy}/m)^{\frac{a-n}{2}}}{\pi^{n/2 -1}2^{(a+n)/2-1}\Gamma(a/2)} 
J_{\frac{a-n}{2}}(m \tau_{xy})
\ee
which obeys
\be
(\Box_x + m^2)w^{a+2}=(\Box_y + m^2)w^{a+2}=w^a
\ee
it being a solution of the Klein-Gordon equation when $a=2$:
\be
(\Box + m^2)w^{2}=0.
\ee
The corresponding Riesz integral is
\be
I^a_{(m)} f(x)\equiv \int_{D} f(y) w^a(x,y) dy
\ee
and it obeys relations analogous to the ones we got in the massless case, i.e.
\bea
&& I^a_{(m)} I^b_{(m)} = I^{a+b}_{(m)}\nonumber\\
&& (\Box + m^2)I^{a+2}_{(m)} = I^a_{(m)}
\eea


Stokes theorem can again be applied to the region $D$, with
two arbitrary functions (0-forms) $f$ and $g$,

\begin{displaymath} \label{principal}
\int_{D}{\left[\left(f d * d g + m^2 f g\ dv\right) - \left(d * d f \wedge g + m^2 f g\ dv\right)\right]} = \nonumber
\end{displaymath}
\be
\int_{\partial D}{\left[ f \wedge * d g  -  * d f \wedge g \right]}
\ee
where $dv=\sqrt{|\eta|}\ d^n x$ is the volume form in $D$.

Let us now take $f$ a solution of the Klein-Gordon equation and $g$ a particular 
combination of funtions, namely
\be 
f=\phi 
\ee
so that
\be
 \left(\frac{1}{\sqrt{|\eta|}}\ d*d  +m^2 \right)\phi = (\Box + m^2) 
\phi = F(x)
\ee
and
\be
g^{(\l)}=\frac{\left(\frac{\t}{m}  \right)^{\l}J_{\l}(m\t)}
{\p^{\frac{n-2}{2}}2^{\l+n-1}\Gamma(\frac{2\l+n}{2}) }
\ee
which obeys
\be
 \left( \frac{1}{\sqrt{|\eta|}}\ d*d +m^2\right) g^{(\l)} = 
(\Box + m^2) g^{(\l)} =g^{(\l - 1)} 
\ee
where $\t=d_{PQ}$ is the geodesic distance from $P\equiv({\bar{x}^\m})$ to another point $Q\equiv({x^\m})$ in the domain $D$, and $J_{\nu}$ is the Bessel function $J$ of order $\nu$.

Using again the analytic continuation idea we get
\bea\label{rieszpropa}
\phi(\bar{x})&=&\lim_{\l\to\frac{2-n}{2}}\frac{1}{\p^{\frac{n-2}{2}}2^{\l+n-1}
\Gamma(\frac{4\l+n}{2}) }\left[\ \int_{D}{dt\ d^{n-1}\vec{x} 
\ F(t,\vec{x})\left(\frac{\t}{m} \right)^{\l} J_{\l}(m\t ) }\right. \nonumber \\
&& +\left.\int_{N_+}{\left(\phi\ *d \left[\left(\frac{\t}{m}  
\right)^{\l}J_{\l}(m\t)\right]  -  *d 
\phi\left(\frac{\t}{m}  \right)^{\l}J_{\l}(m\t) \right)}\right]
\eea

Let specialize to the three dimensional case (without sources), with given data 
on the surface $N_+$. It is convenient to put the integral in terms of the proper time $\t$, then the result is
\bea
\phi^{(3)}_m (\bar{x})&=&\sqrt{\frac{m}{8\p}}\lim_{\l\to\frac{-1}{2}}\int_{0}^{2\p}{d\th}\int_{0}^{\sqrt{\bar{x}^{2}}}{d\t \frac{\t^2-\bar{x}^2}{2(\bar{r}\cos({\bar{\th}-\th})-\bar{t})}   \Big{[}  \left( \l\t^{\l-1}J_{\l}(m\t) \right. \nonumber   }\\
&&\left. \left. +m\t^{\l}J'_{\l}(m\t)  \right)\psi(\t,\theta)     - \t^{\l}J_{\l}(m\t)\frac{\partial\psi(\t,\theta)}{\partial \t}  \right]   
\eea
where $\psi$ is the data known over $N_+$. If we perform an integration by parts we obtain an explicit kernel,
\bea
\phi^{(3)}_m(\bar{x})&=&\sqrt{\frac{m}{8\p}}\lim_{\l\to\frac{-1}{2}}\int_{0}^{2\p}{d\th}\int_{0}^{\sqrt{\bar{x}^{2}}}{d\t \frac{1}{(\bar{r}\cos({\bar{\th}}-\th)-
\bar{t})}\left[ \left( (\l+1)\t^2-\l \bar{x}^2 \right)\t^{\l-1}J_{\l}(m\t) +\right.} \nonumber \\
&& \left. m(\t^2-\bar{x}^2)\t^{\l}J'_{\l}(m\t) \right]\psi(\t,\th)
\eea
Please note that in the $m\to 0$ limit we recover the massless result.

\section{Scalar Fields in Milne space}
Let us now consider a scalar field in the interior of the light cone of flat $n+2$-dimensional
space, $C_{+}$, with 
prescribed values on the cone itself, $N_{+}$
\be
S=\int \sqrt{g}d^{n+2}x \frac{1}{2}\big[g^{\m\n}\pd_{\m}\Phi\pd_{\n}\Phi-m^2\Phi^2\big]
\ee
where $g_{\m\n}$ is the flat Minkowski metric and 
\be
\Phi|_{N_{+}}(t=r,\vec{n})=\psi(u,\vec{n})
\ee
Other spins can be treated similarly. All calculations will turn out to be well-defined and
 finite; no ambiguities will arise in the boundary action, whose formula is universal,
for massless as well as massive fields.
\par
Stokes theorem guarantees that the action on shell can be written as
\bea\label{boundary}
&&S_{N_{+}}=-2^{-(2+n)}\int_0^{\infty}du\, u^n\int_{S_n}\sqrt{g_{s}}d\eta_1\wedge\ldots \wedge 
d\eta_n
\psi(u,\vec{\eta})
\frac{d}{du}\psi(u,\vec{\eta})=\nonumber\\
&&-2^{-(n+2)}\int_0^{\infty} du u^n \frac{d}{du}S_{\psi^2}(u)
\eea
($g_s$ is the determinant of the metric on the sphere $S_n$) and 
where the  {\em spherical mean} of the boundary field $\psi^2$ is defined by averaging
 over  the unit sphere (cf. \cite{John})
\be
S_{\psi^2}(u)\equiv\frac{1}{2}\int_{S_n}\sqrt{g_{s}}\,d\eta_1\wedge\ldots \wedge d\eta_n
\,\psi(u,\vec{\eta})^2
\ee
Incidentally, one could think that this whole procedure is not coordinate 
invariant, because it is defined on a null
surface which enjoys a degenerate metric, but actually Stokes'theorem can be formulated
 entirely in terms of forms.
The action can be rewritten as
\be
S=\frac{1}{2 }\int_M \big(d\Phi \wedge *d\Phi\big)=\frac{1}{2 }\int \big(d[
\Phi \wedge *d\Phi]-\Phi\wedge d*d\Phi\big)
\ee
with
\be
d*d\Phi=\nabla_{\a}\nabla^{\a}\Phi d(vol)
\ee
giving on shell
\be
S_{N_{+}}=\frac{1}{2 }\int_{\pd M} \Phi\nabla^{\rho}\Phi \sqrt{g}\epsilon_{\rho\mu_1\ldots
 \mu_{n+1}}dx^{\m_1}\wedge\ldots\wedge dx^{\m_{n+1}}
\ee
which indeed reproduces (\ref{boundary}).
\par
It can be said that the induced determinant is not the determinant of the induced metric
(which indeed vanishes).
\par
Let us define in general scale transformations in  presence of a gravitational field from
Weyl transformations of the metric

\be
g_{\m\n}\rightarrow \Omega^2 g_{\m\n}
\ee
This is the proper generalization of a scale transformation

\be
x\rightarrow \Omega x
\ee
when the metric is not constant.
\par
The naive scale dimension of the field is {\em defined} so that the kinetic energy remain 
invariant, i.e.
\be
\phi\rightarrow \Omega^{- n/2}\phi
\ee

With this set of rules, the complete boundary action
\be
S_{N_{+}}=\frac{1}{2 (n+1)!}\int_{\pd M} \Phi\nabla_{\sigma}g^{\sigma\rho}
\Phi \sqrt{g}\epsilon_{\rho\mu_1\ldots
 \mu_{n+1}}dx^{\m_1}\wedge\ldots\wedge dx^{\m_{n+1}}
\ee
stays invariant as well, provided  boundary fields are assigned the same 
scale dimension as the bulk fields.
\par

\subsection{Point Splitting}
It is possible to bound the integration over the variable $u$ with an upper limit, say $U$.
Assuming that the prescribed boundary values for the field are analytic in this same variable
\be
\psi(u,\vec{\eta})\equiv\sum_{n=0}^{\infty}\psi_n(\vec{\eta})u^n,
\ee
this yields
\be
S_{N_{+}}=-2^{-(n+2)}\int_0^1 d\eta_1\ldots d\eta_n\frac{1}{\sqrt{1-\vec{\eta}^2}}\sum_{n_1,n_2=0}
\frac{U^{n+n_1+n_2}}{n+n_1+n_2}\psi_{n_1}(\vec{\eta})\psi_{n_2}(\vec{\eta})
\ee

We can always introduce point splitting through Riesz'delta function, defined as
\be
\d^{R}_{\e}(\vec{\eta}-\vec{\eta}^{\prime})\equiv\triangle 
\frac{1}{|\vec{\eta}-\vec{\eta}^{\prime}|^{n-2-\e}}=
\frac{\e(n-2-\e)}{(n-2)V_{n-1} |\vec{\eta}-\vec{\eta}^{\prime}|^{n-\e}}
\ee
where $V_{n-1}$ is the volume of the unit sphere $S_{n-1}$. This is
 such that 
\be
lim_{\e\rightarrow 0}=\d^{R}_{\e}(\vec{\eta}-\vec{\eta}^{\prime})=
\d(\vec{\eta}-\vec{\eta}^{\prime})
\ee
The result of the point split is
(with the $lim_{\e\rightarrow 0}$ understood),
\be
S_{N_{+}}=-2^{-(n+2)}\int_0^1 d\vec{\eta}d\vec{\eta}^{\prime}\frac{1}{\sqrt{1-\vec{\eta}^2}}
\sum_{n_1,n_2=0}
\frac{U^{n+n_1+n_2}}{n+n_1+n_2}\psi_{n_1}(\vec{\eta})\psi_{n_2}(\vec{\eta^{\prime}})
\frac{\e(n-2-\e)}{(n-2)V_{n-1} |\vec{\eta}-\vec{\eta}^{\prime}|^{n-\e}}
\ee
It is clearly possible to choose
\be
U^{n+n_1+n_2}\e(n-2-\e)=1
\ee
Nothing prevents us from using a different limit in each term in the sum, i.e.,
\be
\e_{n_1,n_2}\sim\frac{2}{n-2}U^{-(n+n_1+n_2)}
\ee
(the error in the approximation will then be different for each value of $(n_1,n_2)$).
This yields
\be
S_{N_{+}}=-2^{-(n+2)}\int_0^1 d\vec{\eta}d\vec{\eta}^{\prime}\frac{1}{\sqrt{1-\vec{\eta}^2}}
\sum_{n_1,n_2=0}\frac{1}
{n+n_1+n_2}\psi_{n_1}(\vec{\eta})\psi_{n_2}(\vec{\eta^{\prime}})
\frac{1}{(n-2)V_{n-1} |\vec{\eta}-\vec{\eta}^{\prime}|^{n}}
\ee
\subsection{Boundary-boundary Propagators for massless scalar fields}
Let us introduce the definition in terms of the Riemann-Liouville integral (cf. equation 
(\ref{rl})
\be
\phi_{\l}\equiv I^{2\l +n -2}\phi
\ee
and using the fact that $I^0\phi=\phi$, we can write $\phi_{1- n/2}=\phi$, 
assumed to be on shell.
\par
We have
\be
\Box I^a = I^{a-2}
\ee
which translates into
\be
\Box \phi_{\l}=\phi_{\l-1}
\ee
(All $I^{a}\phi$ vanish when $a$ is a negative integer owing to the fact that the 
constant prefactor then involves a  Gamma function of a negative integer in the denominator.
This forces upon us 
\be
\phi_{\l}=0
\ee
for $\l = 1- n/2-\mathbb{Z}^{+}$)
\par

Using  the boundary limit \footnote{This limit is always delicate, because {\em sensu stricto} it is always a delta
function. Any finite result necessarily involves some sort of smearing of this delta function.
}
of Riesz' boundary-bulk propagator (\ref{propa}), the boundary action for a four-dimensional 
massless scalar field is 

\be\label{ba}
S^{\lambda}_{N_{+}} = 
- \int dt dt'\int_{S_2 (t)} t^2 d\Omega \int_{S_2 (t^{\prime})} (t^{\prime})^2 d\Omega' \ \ \psi(t,\Omega) \  
D_{\l}(t,t^{\prime},\Omega,\Omega^{\prime}) \ \psi(t',\Omega')-S^{\l}_{bulk}
\ee
where the bulk contribution is due to our analytic regularization:
\be
S^{\l}_{bulk}=\int \sqrt{g}d^{n+2}x \frac{1}{2}\Phi_{\l}\Phi_{\l-1}
\ee
and vanishes when $\l$ takes on its physical value.
\par
The propagator is proportional to the differentiated kernel, namely
\bea
&&D_{\l}(t,t^{\prime},\Omega,\Omega^{\prime})\equiv \frac{1}{t^2 (t^{\prime})^2}(\pd_t+\pd_r)
K|_{N_{+}}=
\frac{2}{t^2 (t^{\prime})^2}\pd_{u}K|_{N_{+}} = -\frac{2\l(1+\l)^2}{\pi}\frac{1}{t^{\prime}}
t^{\l-3}
\nonumber\\
&&\left[t(1-\l^2)
-2t^{\prime}\hh \right]^{\l-2} \left[ t^2 (1-\l^2)(1-\l) -2tt'\hh(1-\l^2)
+2(t^{\prime})^2\hh^2 \right]
\eea
When $\l\rightarrow -1 $ it vanishes owing to the $(1+\l)^2$ factor in front. 
What happens when computing the solution is that if the integrations in the action 
(\ref{ba}) are performed in the correct order, that is, for generic $\l$ before taking the 
limit, $\l\rightarrow -1 $
a pole in $\frac{1}{(1+\l)^2}$ appears which 
cancels the above zero in such a way the the ensuing limit is smooth. 
\par
The situation is not unlike the one with evanescent operators when analyzing
anomalies using dimensional regularization.
It is essential to keep this point in mind when use is made of Goursat propagators.
\par
The scale dimension (corresponding to $x\rightarrow \Omega x$) of the scalar field is 
$d(\psi)=-1$. The propagator scales as $ D\rightarrow \Omega^{2\l-4} D$
which implies for the boundary action
\be
 S_{N_{+}}\rightarrow \Omega^{2\l+2} S_{N_{+}}
\ee

\subsection{The regularized boundary and the infinite curvature limit}
Let us now imagine that we regulate Milne's space and we {\em define} it as the interior
of the hyperboloid 
\be
uv=\frac{1}{m^2}
\ee
The characteristic problem is now replaced by a Cauchy problem, which will reduce 
to the corresponding
Goursat problem when $\m\rightarrow\infty$ if the initial value of the derivative is 
chosen in an adequate way. In doing so we are
losing one of the most important properties of our problem, namely, its conformally 
invariant character.
\par  
Incidentally, the induced metric on all hyperboloids is what could be called 
{\em euclidean anti-de Sitter}, (cf. \cite{Alvarez}) 
with isometry group $SO(1,3)$ \footnote{This means that in many respects EAdS 
is more similar to de Sitter than to AdS.}
which in our conventions has 
all coordinates
spacelike
\be
ds^2_{ind}=-\frac{1}{1+m^2 r^2}dr^2-r^2d\Omega^2
\ee
When the parameter $m\rightarrow\infty$ the euclidean $EAdS$ metric degenerates in the 
metric of the light cone $N_{+}$.
It can then be said in this  sense that the lightcone is the infinite curvature limit of $EAdS$.
Precisely in this limit the isometry group is inherited from the adequate form of the 
anti de Sitter group namely
$SO(1,3)$, the four-dimensional Lorentz group,
as we have already seen.
\par
Please remember now that although the boundary of a boundary vanishes, the {\em conformal boundary} of a boundary
(whether conformal or not) does not have to vanish, as we are witnessing now.

\par

We can expect this approximation (that is, the metric of $EAdS$ to look like a light cone 
$N_{+}$) to be valid for
length scales much larger than 
the one defined by the curvature inverse, i.e. it is a low energy approximation, 
valid for $E<< l^{-1}$.
\par

For the light cone itself horospheric coordinates are somewhat similar to the flat coordinates
using to represent the sphere in stereographic projection, {\em em except} that we
now have translational invariance in the $y$-coordinates..
The (singular) boundary-boundary propagator we get from $EAdS$ 
in this way (when $l\equiv \e\rightarrow 0$) is:
\be
\Delta_{b-b}\equiv\frac{\e^{n}\Gamma(n)}{\pi^{n/2}\Gamma(\frac{n}{2})}\,\frac{z^{n}}{|\vec{y}-
\vec{y}^{\prime}|^{n}}
\ee

This propagator is translationally invariant.

\section{The effect of nontrivial gravitational fields}
The most interesting situation from the physical point of view is when there is a nontrivial 
gravitational
field present. Mathematically this means that 
the metric
\be
ds^2(y,\rho)\equiv h_{ij}(y,\rho)dy^i dy^j
\ee
is  not flat. It would be simpler if the metric {\em at the finite boundary} were 
still flat, i.e.
\be
h_{ij}(y,\rho=0)=- \d_{ij}
\ee
The uniqueness of Goursat's problem for Einstein's equations, however, implies that in this 
case
the whole $n+2$-dimensional manifold has to be flat. 
\par
This means that we have to allow 
for a nontrivial metric at the boundary,
\be
h_{ij}(y,\rho=0))\equiv g_{ij}(y)
\ee
where the metric depends only on the combinations (\ref{cartesian})
\be
y^i\equiv\frac{x^i}{x_{-}}
\ee
Assuming analiticity, we can expand all fields in the neighborhood of the boundary,
located at $\rho=0$:
\be
h_{ij}(y,\rho)=g_{ij}(y)+\rho h^{(1)}_{ij}(y)+\ldots
\ee
In cartesian coordinates, $x^{\m}$, the boundary $\rho=0$ coincides with the light cone $x^2=0$.
The fact that the coordinate $\rho$ has to be positive means that the whole manifold  is still
in cartesian coordinates the interior of the forward light cone of the origin: a sort of
curved Milne space.
\par
The metric reads
\be
ds^2=\eta_{\m\n}dx^{\m}dx^{\n}+\frac{x_{-}^2}{l^2} (g_{ij}+\d_{ij})d y^i d y^j + 
\frac{x^2}{l^2} h^{(1)}_{ij}(y^i)dy^i dy^j + \ldots
\ee
There are then two modifications to the flat cone picture worked out in previous sections of
the paper: the first one is proportional to   $(g_{ij}+\d_{ij})$ (that is,
to the {\em initial conditions of the gravitational Goursat problem}), and the
second one is proportional to $x^2$; that is, to the distance to the boundary.
\par
 Even the first term implies all sort of non-diagonal terms in the metric in cartesian coordinates
of the type $d x^i dt$, $d x_{n+1} dt$, $dx^i dx_{n+1}$ as well as modifications of the old 
diagonal terms. 
\par
We are reaching the point in which the use of canonical coordinates is more or less
unavoidable.

\subsection{Canonical coordinates}
Canonical coordinates, although better suited for generalizations are not too easy to visualize.
For example, the light cone of the origin reads 
\be
N_{+}(0)\equiv\{\rho T^2=0\}
\ee
whereas the past light cone of the point $P$, with barred coordinates takes the 
complicated expression:
\be
N_{-}(P)\equiv\{(1-\frac{T}{\bar{T}})\rho+(1-\frac{\bar{T}}{T})\bar{\rho}+
\frac{(\vec{y}-\vec{\bar{y}})^2}{l^2}=0\}\cup \{T T^{\prime}=0\}
\ee

and the Huyghens surface is even more cumbersome: 
\be
{\cal H}\equiv\{(1-\frac{\bar{T}}{T})\bar{\rho}+
\frac{(\vec{y}-\vec{\bar{y}})^2}{l^2}=0\}\cup \{T T^{\prime}=0\}
\ee
\par
In conclusion, even when working with canonical coordinates (which we will do from 
now on), it is convenient to keep
in mind the geometrical setup corresponding to cartesian coordinates, i.e., 
the boundary as a light cone of the origin in some 
extended manifold.
\par

\subsection{Brown-York quasilocal energy}
It is interesting to characterize the gravitational field through the Brown-York (BY) quasilocal
 energy (cf.\cite{Brown}\cite{Liu}) which is defined for a large set of boundary conditions.
\par
To begin with, a definition of {\em time evolution} is necessary. This is equivalent to a 
foliation by a family of spacelike hypersurfaces, which will be denoted by $\Sigma_{\perp}$
for reasons that shall become apparent in a moment. 
\par
\bi
\item
It could seem that the simplest of those is 
\be
T=constant
\ee
This are fine, except that its elements are null surfaces. Some limiting process, akin to the one
worked out in \cite{Brown1} is then necessary.
\par
\item
On the other hand, constant cartesian time (\ref{cartesian})
\be
t=constant=\frac{T}{2}(1+\rho+\frac{\vec{y}^2}{l^2})
\ee
enjoy a normal vector proportional to
\be
\frac{\pd}{\pd T}+\frac{1}{T}(1-\rho+\frac{\vec{y}^2}{l^2})\frac{\pd}{\pd \rho}+\frac{1}{T}
h^{ij}y_j \frac{\pd}{\pd y^i}
\ee
The problem with that is that it is now necessary to regularize the boundary of 
the extended $n+2$ spacetime, by defining a new function
\be
f(T,\rho,y)=0
\ee
say. In order to aply
 the BY formalism, it is exceedingly convenient to choose a boundary which intersects
the foliation orthogonally. 
 The corresponding differential equation is in this case quite complicated, and depend on the $y$-
coordinates.
\item
This leads to our next  foliation by spacelike hyperboloids
\be
t^2- r^2=\bar{t}^2
\ee
(which reduce to the cone itself when $\bar {t}\rightarrow 0$).
\par
In canonical coordinates this foliation reads
\be
\rho T^2 = \bar{t}^2
\ee
\par
Please note that this corresponds to surfaces of constant norm of the CKV (\ref{ckv}).
\par
The unit normal vector in this case is given by
\be
\frac{T}{\sqrt{\bar{t}}}\frac{\pd}{\pd T}
\ee
so that the boundary of the spacetime is orthogonal to the foliation as long as
\be
\frac{\pd}{\pd T}f=0
\ee
In cartesian coordinates this means that
\be
\pd_{-}f=0
\ee
\item Perhaps the simplest possibility is the one we formerly used in \cite{Alvarez}, that is,
\be
\rho=\e
\ee
This leads to an energy
\be
E\equiv \frac{1}{\kappa^2_{n+2}}\int_{B_n} d(vol)_n (K-K_0)
\ee
where $K$ is the trace of the extrinsic curvature of the imbedding $B_n\hookrightarrow N_{\e}$
and $K_0$ the same quantity for the imbedding $B_n\hookrightarrow \mathbb{R}_n$
\par
The quasilocal energy is then defined in the $n$-dimensional surface 
$B_n\equiv \pd M\cap \Sigma_{\perp}$, the intersection of the boundary of our spacetime, 
$\pd M$, with a leave in the chosen foliation say, $\Sigma_{\perp}$, which is a codimension
two submanifold,whose metric is
\be
ds^2=\frac{L^2}{\epsilon l^2}h_{ij}dx^i dx^j
\ee
Using the Ricci-flatness condition (that is, Einstein's equations),
it has been shown in \cite{Alvarez} that the energy can be expressed as:
\be
E=-\frac{L^{n-1}}{\kappa^2}\int_{B\cap\Sigma_{\perp}} \frac{1}{l^n\epsilon^{n/2}}\sqrt{h}d^n x 
(-n +\epsilon h^{kl}h^{ \prime}_{kl})
\ee
 The quasilocal 
energy has to be refered to a particular template, which is to be attributed
 the zero of energy. In our case this would mean to substract the energy 
of the flat six dimensional space, and stay with
\be
E=- \frac{1}{\kappa^{2}}\int{    d^{n}x\ \frac{L^{n-1}}{l^{n}
\e^{\frac{n}{2}}}\sqrt{|h|}\e h^{ij}h_{ij}^{\prime}}
\ee
which is such that its $\e$-independent part is 
 proportional to $E_4 + W_4$ with non-zero coefficient where  $E_4$ is the 
integrand of the four dimensional Euler character and
$W_4$ is the four dimensional quadratic Weyl invariant.

\par
It is indeed remarkable that this is the correct form (up to 
normalization) for the conformal anomaly for conformal invariant matter (we have checked
that this remains true in six dimensions). 
This fact allows for an identification of the central function of the putative CFT, 
namely,
\be
c=\frac{l^{4}}{\kappa_6^2}
\ee
Although at first sight it is natural to take $\e\rightarrow 0$, because in that way we cover
as much space as possible, this is a point in need of clarification.
\par

\ei
\subsection{Perturbative expansion}
Written in canonical coordinates, the wave operator reads
\be
\Box=\frac{l^2}{T^2}\Box_{y}[h]+\frac{4}{T}\frac{\pd}{\pd \rho}\frac{\pd}{\pd T}+
\frac{2n-4}{T^2}\frac{\pd}{\pd \rho}-\frac{4 \rho}{T^2}\frac{\pd^2}{\pd \rho^2}+
\frac{2}{T}\frac{\partial\ ln\sqrt{h}}{\pd \r}\frac{\pd}{\pd T}
-\frac{4\r}{T^2} \frac{\pd\ ln\sqrt{h}}{\partial \r}\frac{\pd}{\pd \r}
\ee
All the difference from the wave operator corresponding to Milne space
\be
 \Box_{x}=\frac{l^2}{T^2}\Box_{y}+\frac{4}{T}\frac{\pd}{\pd \rho}\frac{\pd}{\pd T}+
\frac{2n-4}{T^2}\frac{\pd}{\pd \rho}-\frac{4 \rho}{T^2}\frac{\pd^2}{\pd \rho^2}
\ee
stems from the appearance of $h_{ij}(x,\rho)$ instead of $\d_{ij}$.
The interpretation of the full space as the interior of a cone in flat space is, of course, lost.
However, the fact that $h_{ij}$ is analytic around the boundary, $\rho=0$ means that the full 
wave operator can be written as
\be
\Box=\Box_x +\sum_{n=1}^{\infty}\rho^n D_n
\ee
where $D_n$ are differential operators, whose explicit form is known in terms of the metric
 $h_{ij}$.
It is now a simple matter to solve the scalar wave equation perturbatively, by representing
\be
\Phi\equiv \sum\Phi_n\rho^n
\ee
 We have, for example,
\be\label{iter}
(\Box_x + m^2)\Phi_1=- D_1 \Phi_0
\ee
where $\Phi_0$ is a solution of the Minkowskian equation
\be
(\Box_x + m^2)\Phi_0=0
\ee
The introduction of a nontrivial gravitational background does not present 
then any problems in principle: the equation (\ref{iter}) can be solved by using 
Riesz'propagator
for the Klein-Gordon equation with known second member $- D_1 \Phi_0$ (\ref{rieszpropa}).

\section{Concluding remarks}

In conclusion, we have unveiled a rich conformal structure in the finite conformal boundary of
the Ricci flat family of spacetimes with vanishing cosmological constant 
we have endeavoured to study. The main property
of those spaces from the present point of view is that the Cauchy problem is not well posed,
and one has to solve a Goursat, or characteristic problem instead. Once one does that, there
is a mapping between fields at the finite boundary and fields in the bulk.
\par
Actually, we have got two such propagators; one coming from the Riesz potential, which is
not manifestly translationally invariant in spherical coordinates, 
and another one coming from the infinite curvature
limit of constant curvature spaces, which enjoys this property in horospheric coordinates.
 We have argued for the
physical equivalence of both approaches, but a more explicit treatment is perhaps desirable.
The relationship of this whole mapping with the matrix models of \cite{Banks} is 
intriguing and seems
worth exploring in detail.
\par
The fact that the trace of the Brown-York energy-momentum tensor is proportional (in four
euclidean dimensions) to the conformal anomaly for conformally invariant matter is probably
suggesting a relationship with some conformal field theory, although we have not been
able to compute the precise value of the proportionality constant.
\par
It also remains to study the interplay between the {\em finite boundary} at $\rho=0$
 and $\mathcal{J}^{+}$ ,
 which is
a sort of half an S-matrix, in the 
hope that it will unveil some of the mysteries of holography with vanishing 
cosmological constant.

\section*{Acknowledgments}

This work has been partially supported by the
European Commission (HPRN-CT-200-00148) and FPA2003-04597 (DGI del MCyT, Spain).      
E.A. is grateful to G. Arcioni, J.L.F. Barb\'on and E. Lozano-Tellechea 
for discussions and to Jaume Garriga and Enric 
Verdaguer for useful correspondence. E.A. is also grateful to Ofer Aharony and the other 
members of the Weizmann Institute of Science, where this work was completed, for 
their kind invitation.

\appendix

\section{Conformal isometries of the light cone}

It is quite simple to check that, for example boosts in the $(n+1)$ direction lead to
scale transformations on the sphere $S_n$:
Calling $x\equiv x_{n+1}$ and $\vec{x}_T \equiv (x^1\ldots x^n)$ the boost reads, in terms of
 the rapidity $\chi$ (we shall employ in this section the notation $x^0\equiv t$).
\bea
&&t^{\prime}=t\, \cosh{\chi}+x\, \sinh{\chi}\nonumber\\
&&x^{\prime}= t\, \sinh{\chi}+x\, \cosh{\chi}
\eea
which on the cone $t=r$ (so that {\em a fortiori}, $t^{\prime}=r^{\prime}$), yields
\bea
&&r^{\prime}=r\, \cosh{\chi}+x\, \sinh{\chi}\nonumber\\
&&x^{\prime}= r\, \sinh{\chi}+x\, \cosh{\chi}
\eea
This gives ($n\equiv\frac{x}{r}=\sqrt{1-\vec{\eta}^2}$ with $\vec{\eta}\equiv\frac{\vec{x}_T}{r}$)
\be
\frac{r^{\prime}}{r}= n \, \sinh{\chi}+ \cosh{\chi}
\ee
which is enough to determine the transformation of the metric on $N^{+}$ through:
\be
\vec{\eta}^{\prime}=\frac{1}{\cosh{\chi}+\sinh{\chi}\sqrt{1-\vec{\eta}^2}}\vec{\eta}
\ee

\be
ds_{+}^2=t^2 \big[\frac{(\vec{\eta} d\vec{\eta})^2}{1-\vec{\eta}^2}+ d\vec{\eta}^2\big]=
t^2\big(\d_{ij}+\frac{\eta_i\eta_j}{1-\vec{\eta}^2}\big)d\eta_i d\eta_j
\ee 
namely (taking into account the transformation of the prefactor $t^2$)
\be
(ds^{\prime}_{+})^2=(\cosh{\chi}+n\, \sinh{\chi})^2 \frac{1}{(\cosh{\chi}+ 
n\, \sinh{\chi})^2}ds_{+}^2
\ee
\par
Note that precisely
\be
\sqrt{g_{+}^{\prime}}=\sqrt{g_{+}}(\cosh{\chi}+n\, \sinh{\chi})^{-n}
\ee
\vspace{1cm}
This relationship is actually quite general: by noticing that the equation of the sphere $S_n$
\be
\sum_1^n (n^i_T)^2+n^2 = 1
\ee
can be rewritten as

\be
\sum_1^n (x_i)^2+ x_{n+1}^2 = x_0^2
\ee
It is now clear that a Lorentz transformation of $O(1,n+1)$ acts projectively on the sphere; from
\be
(x^{\m})^{\prime}= L^{\m}\,_{\n}x^{\n}
\ee
we get for $a,b=1\ldots (n+1)$
\be
(n^a)^{\prime}=\frac{L^a\,_0+L^a\,_b n^b}{L^0\,_0+L^0\,_b n^b}
\ee
i.e., the Moebius group of conformal transformations of the sphere, which are known to exhaust
all conformal motions of $S_n$. Only when $L^0\,_0=1$ and $L^0\,_b=0$
(that is, for pure $O(n+1)$ rotations) do we get an isometry.
\par
The general Moebius transformation in terms of the independent variables reads:
\be
\eta_i^{\prime}=\frac{L^i\,_0+L^i\,_{n+1}\sqrt{1-\vec{\eta}^2}+L^i\,_l\eta^j}{L^0\,_0+
L^0\,_{n+1}\sqrt{1-\vec{\eta}^2}+L^0\,_j\eta^j}
\ee

\par
Physics on the cone is quite simple: time evolution is equivalent to a scale transformation
on the radius of the sphere.
\par
In cartesian coordinates the degenerate metric on the cone reads
\be
ds^2=(\frac{x^i x^j}{r^2}-\d_{ij})dx^i dx^j
\ee

\subsection{Three and four-dimensions}
We shall give in the main body of the paper explicit examples of propagators in three and
four dimensions. Let us begin here by analyzing conformal motions in detail for those examples.
\par
In $d=1+2$ dimensions the metric on the cone reads
\be
ds^2= t^2(1+\frac{\eta^2}{1-\eta^2})d\eta^2 = t^2 d\theta^2
\ee
with the identification $\eta\equiv \sin{\theta}$.
\par
 A boost such as
\begin{displaymath}
\mathbf{B}=
\left(\begin{array}{ccc}
\cosh{\chi}&\sinh{\chi}&0\\
\sinh{\chi}&\cosh{\chi}&0\\
0&0&1
\end{array}\right)
\end{displaymath}
induces the transformation
\be
\eta^{\prime}=\frac{\sinh{\chi}+\eta \cosh{\chi}}{\cosh{\chi}+\eta\sinh{\chi}}
\ee
and
\be
\frac{t^{\prime}}{t}=\frac{r^{\prime}}{r}= \cos{\chi}+\eta\sinh{\chi}
\ee
namely,
\be
d\theta^{\prime}=\frac{d\theta}{\cosh{\chi}+\sinh{\chi}\sin{\theta}}
\ee
\par
In $d=1+3$ dimensions the metric on the cone is usefully written in complex coordinates
through stereographic projection
\be
z\equiv \frac{n_1 + i n_2}{1-n_3}
\ee
A boost such as

\begin{displaymath}
\mathbf{B}=
\left(\begin{array}{cccc}
\cosh{\chi}&0&0&\sinh{\chi}\\
0&1&0&0\\
0&0&1&0\\
\sinh{\chi}&0&0&\cosh{\chi}\\
\end{array}\right)
\end{displaymath}
induces the transformation
\bea
&&n_1^{\prime}=\frac{n_1}{\cosh{\chi}+n_3\sinh{\chi}}\nonumber\\
&&n_2^{\prime}=\frac{n_2}{\cosh{\chi}+n_3\sinh{\chi}}\nonumber\\
&&n_3^{\prime}=\frac{\sinh{\chi}+n_3\cosh{\chi}}{\cosh{\chi}+n_3\sinh{\chi}}
\eea
and
\be
\frac{t^{\prime}}{t}=\frac{r^{\prime}}{r}= \cosh{\chi}+n_3\sinh{\chi}
\ee
that is
\be
z^{\prime}=\frac{1}{\cosh{\chi}-\sinh{\chi}}z=(\cosh{\chi}+\sinh{\chi}) z
\ee
as well as
\be
\frac{t^{\prime}}{t}=\frac{r^{\prime}}{r}= \frac{\cosh{\chi}+\sinh{\chi}+|z|^2 (\cosh{\chi}-
\sinh{\chi})}{1+|z|^2}
\ee
which acts as a conformal transformation on the metric of the Riemann sphere $S^2$
\be
ds^2=\frac{4 dz d\bar{z}}{(1+|z|^2)^2}
\ee
and is compensated on the cone $t=r$ by the transformation of $t^2$.
A well-known theorem (cf., for example, \cite{Penrose}) guarantees that all Lorentz 
transformations can be represented as elements of $SL(2,\mathbb{C})$:
\be
z^{\prime}=\frac{a z + b}{c z + d}
\ee
where $a,b,c,d \in \mathbb{C}$ and $ad-bc=1$.
\section{Fractional Derivatives and The Riemann-Liouville Integral}
Fractional derivatives can be defined from
\be
D^{-1}f\equiv\int_0^x f(t)dt
\ee
and
\be
D^{-n}f\equiv\frac{1}{(n-1)!}\int_0^x f(t) (x-t)^{n-1} dt
\ee
so that
\bea
&&D^{-1} D^{-n}f=\int_0^x dt \frac{1}{(n-1)!}\int_0^t du f(u)(t-u)^{n-1}=
\int_0^x du\frac{1}{(n-1)!}\int_u^x dt f(u) (t-u)^{n-1}=\nonumber\\
&&\frac{1}{n!}\int_0^x du f(u) (t-u)^n\bigg|^x_u
\equiv D^{-(n+1)}f
\eea
The Riemann-Liouville integral is the inmediate generalization of this, namely,
\be
D^{-\lambda}f(x)\equiv\frac{1}{\Gamma(\lambda)}\int_a^x dt f(t) (x-t)^{\lambda-1}
\ee
It is now easy to check the basic property that
\be
D^0 f(x)\equiv \lim_{\lambda\rightarrow 0}D^{\lambda}f=\lim_{\lambda\rightarrow 0}\lambda\bigg(
\int_a^x dt (x-t)^{\lambda}\frac{f^{\prime}}{\lambda}- f(t)\frac{(x-t)^{\lambda}}{\lambda}\bigg|_a^x\bigg)=f(x)
\ee

The n-dimensional generalization is straightforward in the euclidean case, and a 
litle bit less so in 
the lorentzian situation, which is the one considered by Riesz.

\section{Timelike boundary versus null boundary.}

In the well-known example of Maldacena's AdS/CFT (\cite{Maldacena}\cite{Witten}) 
the way it works is that data are given on
the conformal (Penrose) boundary, which is timelike. In global coordinates the space reads
\be
ds^2=\frac{l^2}{cos^2(\rho)}(d\tau^2-d\rho^2-sin^2(\rho)d\Omega_{d-2}^2)
\ee
where $0\leq \rho < \pi/2$ and $0\leq \tau < 2\pi$,
and the boundary is located at $\rho=\pi/2$, a timelike surface indeed.
The space as such has closed timelike curves, a fact which disappears when
 the covering space is considered (namely $-\infty < \tau < \infty$). In this case, the domain 
of dependence of data on the boundary encloses the full space.
\par
Cauchy's problem with data on a timelike surface does not have a solution in general.
Goursat \cite{Goursat} has studied some particular examples.

\begin{figure}[!ht] 
\begin{center} 
\leavevmode 
\epsfxsize= 5cm

\epsffile{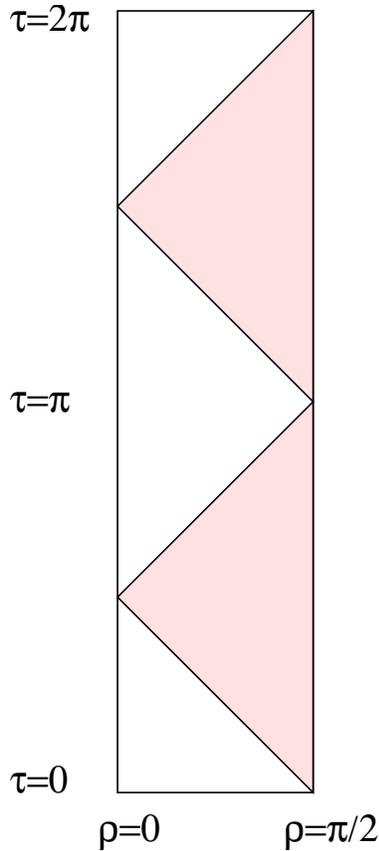} 

\caption{{\em Horospheric coordinates only cover half of the whole anti de Sitter space (
namely the portion which is not filled).}}
\label{fig:uno} 
\end{center} 
\end{figure}
\par

In the AdS case however the solution of the related riemannian Laplace 
problem, 
that is, the operator obtained by replacing all minus signs by pluses in the metric, 
gives results for the propagator which are related by analytical continuation to the ones 
obtained with lorentzian signature (except for some subtleties related to 
the presence of normalizable modes (\cite{Witten}).
\par

This is not true anymore for  the  Goursat's characteristic problem. Let us consider, the
three-dimensional case to be specific.
The analytical continuation of the hyperboloid
\be
H_{\e}\equiv t^2 - x^2-y^2 = \e^2
\ee
is the sphere
\be
S_{\e}\equiv t^2 + x^2 + y^2 = \e^2
\ee
The solution to the exterior Dirichlet problem on the sphere $S_{\e}$
for Laplace's equation is given by the Poisson integral (cf. \cite{Courant})
which expressed in three-dimensional polar coordinates reads
\be
\Phi (R,\theta,\phi)=\frac{\e (R^2-\e^2)}{4\pi}\int_0^{2\pi}d\phi^{\prime} \int_0^{\pi}
d\theta^{\prime}
\frac{f(\theta^{\prime},\phi^{\prime}) sin\,\theta^{\prime} }{(\e^2+R^2-2\e R cos\,\gamma)^{3/2}}
\ee
where $f(\theta,\phi)$ is the Dirichlet datum on $S_{\e}$ and
\be
cos\,\gamma\equiv cos\,\theta\, cos\,\theta^{\prime}sin\,\theta\, sin\,\theta^{\prime}\,cos\,
(\phi-\phi^{\prime})
\ee
It will prove instructive to consider some examples in detail. In particular, the solution 
which reduces to
\be
t^2\equiv \e^2 cos^2\,\theta
\ee
is easily found \footnote{Using, for example, that in terms of Legendre polynomials,
\be
x^2=\frac{2 }{3}P_2 (x) +\frac{1}{3}P_0(x)
\ee
}
to be:
\be
\Phi_E=\frac{\e^2}{3}+\frac{2 t^2-x^2-y^2}{3}
\ee
The corresponding analytic continuation
\be
\Phi_L=\frac{ \e^2}{3}+\frac{2 t^2+x^2+y^2}{3}
\ee
is indeed a solution of the wave equation with Lorentzian signature which reduces
on $H_{\e}$ to $t^2$.
The problem is that from the point of view of $H_{\e}$ $t^2$ and $r^2$ are the same thing, which
is not true from the point of view of $S_{\e}$. 
\par
For example, the solution of Laplace's equation which reduces on $S_{\e}$ to $r^2=x^2+y^2$ is
\be
\Phi_E=\frac{2 \e^2}{3}+\frac{1}{3}(x^2+y^2- 2 t^2)
\ee
which upon analytic continuation reduces to a solution which tends to $-\phi_{char}$ when
$\e\rightarrow 0$
\par
In the same vein, one could consider the solution of Laplace's equation which reduces to
\be
\lambda t^2 + (1-\lambda)(x^2+y^2)
\ee
All of them are equivalent from the point of view of the characteristic problem. From the Laplace
viewpoint, however, they are equivalent to $(1-\lambda)\e^2 + (2\lambda-1)t^2$.
The result is
\be
\Phi_E=\frac{(2-\lambda)\e^2}{3}+\frac{2\lambda-1}{3}(2t^2-x^2-y^2)=
\frac{(2-\lambda)\e^2}{3}+\frac{2\lambda-1}{3}R^2(2-3 sin^2\,\theta).
\ee
(with $R^2\equiv t^2+ x^2+ y^2$), whose analytic continuation is
\be
\Phi_L=\frac{(2-\lambda)\e^2}{3}+\frac{2\lambda-1}{3}(2t^2+x^2+y^2)
\ee
so that the characteristic solution is recovered when $\lambda=1$ only; the analytic
 continuation is inherently ambiguous. 
\par
Incidentally, we have been considering here up to now 
the {\em interior} solution of Laplace's equation,
whereas it would appear more appropiate to consider the {\em exterior} solution instead, which is
easily obtained through an inversion
\be
\Phi_{ext}(R,\theta,\phi)=\frac{\e}{R}\phi_{int}(\frac{\e^2}{R},\theta,\phi).
\ee
that is
\be
\Phi_{ext}=\frac{(2-\lambda)\e^2}{3}+\frac{2\lambda-1}{3}\frac{\e^5}{R^3}(2-3 sin^2\,\theta)
\ee
which does not have the correct analytic continuation.

\section{Goursat versus Cauchy}
It is interesting to consider the Cauchy problem for initial data on the spacelike
hypersurface $N_{\e}(0)$
\be
t^2-r^2=\e^2 = u.v
\ee
which degenerates into the cone $N_{+}(0)$ when $\e\rightarrow 0$.
\par
It could be thought that nothing changes much, but there is one essential point which does.
Namely, the initial function is now
\be
\psi(u)\equiv\phi(u,v=\frac{\e^2}{u})
\ee
Deriving once and remembering that
\be
f(u)\equiv \frac{\pd \phi}{\pd v}(u,v=\frac{\e^2}{u})
\ee
yields in an obvious notation, evaluating all quantities at $N_{\e}$
\be
\pd_u \psi=\phi_u-\frac{\e^2}{u^2}f
\ee
but there is now no restriction on the function f, given the fact that
\be
f_u=\phi_{uv}-\frac{\e^2}{u^2}\phi_{vv}
\ee
conveying that fact that now $\phi_{uv}$ is {\em not} uniquely 
determined in terms of (derivatives of) f.


We ask ourselves about the relation between the Cauchy and the characteristic problem for the wave equation of a scalar field $\phi$, the first posed on the hyperboloid $t^2 -r^2 = \e^2$ with the value of the field and its normal derivative given, and the second on the future light cone $t=r$ with only the field known. \\

It seems that there is some preferred value for the field normal derivative such that 
in the limit $\e\to 0$ the Cauchy problem goes to the characteristic one, that is, the 
derivative is somehow fixed and we only need the field itself. 

In order to check these ideas, let work out one simple example in detail. We have a three 
dimensional solution to the wave equation with prescribed values on the surface 
$t^2 -r^2 = \e^2$.
\be
\phi =\frac{2t}{3}+\frac{\e^3 t}{3 (t^2 -r^2 )^{3/2}}
\ee
which obeys
\begin{eqnarray}
&& \Box\phi =0    \nonumber \\
& & \left.\phi \right|_{t^2 -r^2 =\e^2 }=t      \nonumber  \\
&& \left.\frac{d\phi}{dn}\right|_{t^2-r^2=\e^2}=0    
\end{eqnarray}

When $\e\to 0$ the solution goes to
\be
\hat{\phi}=\frac{2t}{3}
\ee
 which is not the solution to the characteristic problem with the desired boundary value. But we also know this one, which is
\be
\psi=t \qquad \to  \qquad  \psi |_{t^2-r^2=\e^2}=t      
\ee

The normal derivative of the Cauchy problem is $\frac{d}{dn}=\frac{1}{\e}(t\partial_t +r\partial_r )$ and it is not extended to the null surface. Nevertheless, taking the characteristic solution, we can perform the derivative $t\partial_t +r\partial_r$ and then regulate,
\be
\frac{d}{dn}\phi = \frac{1}{\e}(t\partial_t +r\partial_r) \psi = \frac{t}{\e}
\ee

We solve again the Cauchy problem with this prescribed derivative and find easily
\be
\phi =\frac{2t}{3}+\frac{\e^3 t}{(t^2 -r^2 )^{3/2}} +\frac{t}{3}
\left( 1-\frac{\e^3}{(t^2 -r^2 )^{3/2}} \right)
\ee
which satisfies
\begin{eqnarray}
&& \Box\phi =0    \nonumber \\
&& \left.\phi \right|_{t^2 -r^2 =\e^2 }=t      \nonumber  \\
&& \left.\frac{d\phi}{dn}\right|_{t^2-r^2=\e^2}= \frac{t}{\e}  
\end{eqnarray}

This solution is precisely the desired one in the $\e\to 0$ limit,
\be
\phi |_{t^2-r^2=\e^2}= \psi= t
\ee

\section{T-dual Formulation}

It is possible to assume that the coordinates $y^i$ (cf. \ref{cartesian}),
 $1\leq i\leq n$ live in an torus,  
for example
\be\label{TD}
y^i= L \theta^i
\ee
This means that
\be
x^i=\frac{L}{l}x_{-}\theta^i
\ee
That is, the combination $\frac{x^i}{x_{-}}$ must be a periodic function.
\par
On the other hand, if 
\be\label{TDD}
x^i = R \phi^i
\ee
then
\be
y^i = \frac{l R}{x_{-}}\phi^i
\ee
Let us concentrate for example in this last possibility (\ref{TDD}).
The main difference with uncompactified Minkowski space is that now Goursat's problem
on the fundamental domain leads to a solution only in the diamond-like region highlighted in
the figure. In order to get a solution on the whole space, data are needed on all lines
\be
t=\pm r + n R
\ee
where $n\in\mathbb{N}$,
\begin{figure}[!ht] 
\begin{center} 
\leavevmode 
\epsfxsize= 5cm

\epsffile{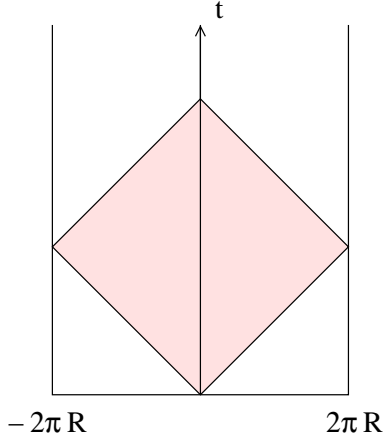} 

\caption{{\em The fundamental diamond with angular coordinates.}}
\label{fig:cinco} 
\end{center} 
\end{figure}
\par
If we interpret the full (n+2)-dimensional space as a string background (that is, we complete 
it with an extra 
internal Ricci-flat compact manifold of dimension $D=8-n$, and we interpret the scale $l$ as the
string scale $l_s$)
then it is known \cite{Giveon} to be equivalent to its T-dual.
\par
This means, if for example  (\ref{TD}) were true, that the gravitational background would read
\be
d\tilde{s}^2=-\frac{ l_s^4}{ L^2 T^2}d\vec{\tilde{\theta}}^2 +\rho dT^2 + TdT d\rho
\ee 
with a nontrivial dilaton as well:
\be
\Phi\equiv -\frac{n}{2}\log{\frac{T^2 L^2}{l_s^2}}
\ee
This space is not flat, so that it cannot be equivalent to a wedge of Minkowski.
Non-constant dilatons, however, are notoriously difficult to analyze.


\end{document}